\documentclass[sigconf]{acmart}
\usepackage[utf8]{inputenc}
\usepackage{paralist}

\usepackage{url}
\makeatletter
\g@addto@macro{\UrlBreaks}{\UrlOrds}
\makeatother

\setcopyright{none}
\renewcommand\footnotetextcopyrightpermission[1]{} %

\title{A Model for Using Physiological Conditions for Proactive Tourist Recommendations}

\author{Rinita Roy}
\affiliation[obeypunctuation=true]{%
	\institution{Technical University of Munich}\\
	\streetaddress{Bolzmannstr. 3}
	\postcode{85748}
	\city{Garching},
	\state{Germany}
}
\email{rinita.roy@tum.de}

\author{Linus W.\,Dietz}
\orcid{orcid.org/0000-0001-6747-3898}
\affiliation[obeypunctuation=true]{%
	\institution{Technical University of Munich}\\
	\streetaddress{Bolzmannstr. 3}
	\postcode{85748}
	\city{Garching},
	\state{Germany}
}
\email{linus.dietz@tum.de}

\begin{abstract}
Mobile proactive tourist recommender systems can support tourists by recommending the best choice depending on different contexts related to herself and the environment.
In this paper, we propose to utilize wearable sensors to gather health information about a tourist and use them for recommending tourist activities.
We discuss a range of wearable devices, sensors to infer physiological conditions of the users, and exemplify the feasibility using a popular self-quantification mobile app.
Our main contribution then comprises a data model to derive relations between the parameters measured by the wearable sensors, such as heart rate, body temperature, blood pressure, and use them to infer the physiological condition of a user.
This model can then be used to derive classes of tourist activities that determine which items should be recommended.
\end{abstract}


\begin{document}
\pagestyle{plain}
\maketitle

\section{Introduction}

Suggesting points of interest or activities to tourists within a city is a popular and challenging recommender systems research problem.
To provide interesting recommendation at the right time, context-awareness is very advantageous~\cite{Woerndl2007}.
Currently, mobile phones are commonly used for eliciting its user's context.
However, in some situations tourists might be uncomfortable pulling out their smart phones due to the fear of them being stolen, or for social reasons as it might be inappropriate to check phones all the time when fellow-travellers are around.
This gives rise to wearable devices, such as smart watches for context modeling.
Wearable devices have embedded sensor technologies and can be used to collect various information about the wearer and the surrounding environment.
Wallace and Press~\cite{Wallace2003} showed new perspectives of looking at wearable objects by involving computer technologies into them.
However, wearable devices of the last ten years such as the Fitbit~Zip\footnote{https://fitbit.com/au/Zip},  Google Glass\footnote{https://developers.google.com/glass/develop/gdk/location-sensors}, Apple Watch\footnote{https://apple.com/lae/watch/}, and the Microsoft HoloLens\footnote{https://microsoft.com/en-IE/hololens} had mixed success on the market and the smart phone is still the most prevalent mobile device.

As a tourist, it is useful to get personalized recommendations about where to visit, what to do next throughout a trip, given that the recommender has sufficient system information about the context of the tourist.
Unlike classical recommender systems, mobile recommendations can also be proactive, meaning that the recommendations are pushed to the user without explicit request~\cite{Woerndl2011}.
This introduces the additional challenge to determine the appropriate timing of the recommendations.
Wearable sensor devices help to gather various data about and around the wearer.
This data can be used to acquire contextual information, and with further processing, the proper conditions can be derived to recommend tourist activities to the wearer.
Our approach was thus guided by the following research questions:

\begin{enumerate}
\item[RQ1:] How can different physiological conditions be inferred using wearable sensors? 
\item[RQ2:] How can these conditions be used to derive preferences for activities? 
\end{enumerate}

In the following section, we survey related work on tourist recommender systems based on context awareness, wearable devices, and sensor data.
We present our data model in \autoref{RA2}, which shows how to select tourist activities based on data from typical sensors found in wearable devices.
Finally, we draw the conclusion and present future work in \autoref{CAndFW}.

\section{Foundations}\label{RW}

We argue that to improve mobile, context-aware, and proactive recommender systems, sensors in wearable devices can be useful by inferring the physiological conditions of a user.
If the sensor data can be interpreted accurately in the  context modeling step, the overall recommendation accuracy of a mobile recommender system can be improved with recommendations that fit well to the context and condition of the user.

\subsection{Context-Aware Recommender Systems}

Throughout this paper, we follow Dey's definition of context: \emph{``Context is any information that can be used to characterise the situation of an entity. An entity is a person, place, or object that is considered relevant to the interaction between a user and an application, including the user and applications themselves~\cite{Dey:2001:UUC:593570.593572}.''}
It is generally believed that if the contextual information of a person, e.g., physical or emotional states, or her surroundings, e.g., location and time, is available, the accuracy of the recommendations increases.
Adomavicius et~al.~\cite{journals/aim/AdomaviciusMRT11} presented a way how context can be used to develop intelligent recommender systems.
Ashley-Dejo et~al.~\cite{AshleyDejo2016ACP} proposed a context-aware proactive recommender system for tourists and determined the accuracy of the recommendations comparing the benefits of using various different contextual information, such as location, weather, or time as opposed to not considering context at all.
In their evaluation the traditional multi-criteria collaborative filtering approach, where no context information was used had the least performance, while the recommendations where all context information was used performed best.

For the recommendation of customized destinations and routes to travelers, the current location is one of the most prevalent context factors.
Meehan et~al.~\cite{conf/percom/MeehanLCM13} highlight the importance of considering more contextual information other than location, such as, weather, time, sentiment, and user preferences.
But not all context factors are equally important as Baltrunas et~al.~\cite{Baltrunas2012} found when they investigated the importance of a context factors for different activities.
For example, TourRec, a recommender system for city trips~\cite{Lass2017} only uses time of the day, weather and previously visited points of interest.
In the model proposed in our paper, we showcase the possibility of using physiological conditions for proactively recommending personalized tourist activities.

\subsection{Wearable Devices}

Wearable devices can support humans with the overload of information in communication and computation in appropriate contexts~\cite{journals/computer/BillinghurstS99}.
A survey of commercially available wearables, research prototypes, and their classification was done by Seneviratne~et~al.~\cite{journals/comsur/SeneviratneHNLK17}, and Godfrey et~al.~\cite{Godfreya2018FromAT} provide an A-Z guide of the key terms required for understanding current wearable technology developments in health care.

Wearable devices are useful for measuring various physiological and biochemical parameters~\cite{10.1088/978-0-7503-1505-0ch1}, which can further be processed to benefit the healthcare domain.
For example, a sensor-equipped ring has been used to provide physiological data of patients and further analyzing it with their profiles by using assistant systems showed that it can help to resolve wrong diagnoses by doctors~\cite{conf/aisi/ShehabIOEE17}.
Clothing or textiles are good collectors of body information as they stay in close contact with the skin~\cite{journals/sensors/StoppaC14}.
Smart textiles are also widely used to noninvasively gather health information~\cite{Coyle2010BIOTEXBiosensingTF} and physical monitoring~\cite{Pandian2008SmartVW}.
Kamisalic et~al.~\cite{journals/sensors/KamisalicFTK18} surveyed how various physiological parameters and activities can be noninvasively measured by sensors incorporated into wrist-worn devices.
For this reason these type of sensors are also useful to monitor athletes' training and health~\cite{10.3389/fphys.2016.00071}.

Despite its potential, wearable devices have not yet become that predominant in the tourism sector.
Tussyadiah~\cite{Tussyadiah2014} identified personal motivations to use wearable devices for travel and tourism.
Atembe~\cite{RolandAtembe2015} compiled the limited use of wearable devices in tourism by presenting some use cases and providing the usage of wearable devices there.
One early work in this area is the one of Vlahakis~et~al.~\cite{conf/iswc/VlahakisKTIS02}, who designed a system that provides interactive, personalized, augmented reality tours in archaeological sites using mobile and wearable computers.

We work towards the usage of wearable devices in the tourism domain by showing a way to derive contextual information by using the data collected by wearable sensors and then utilizing the context to proactively recommend relevant activities during a trip suitable to her physiological conditions.

\subsection{Data Ecosystems}

The basic assumption is that sensors in wearable devices are capable of measuring different health parameters, such as the heart rate or blood pressure, and report this to a centralized application.
Through constant monitoring, the  sensor data can be used to determine different physiological conditions, e.g., to find out how active, tired, or hungry the user is.
In a second step a recommender system can then utilize these conditions as context factors to make recommendations.

\begin{figure}
	\includegraphics[width=\columnwidth]{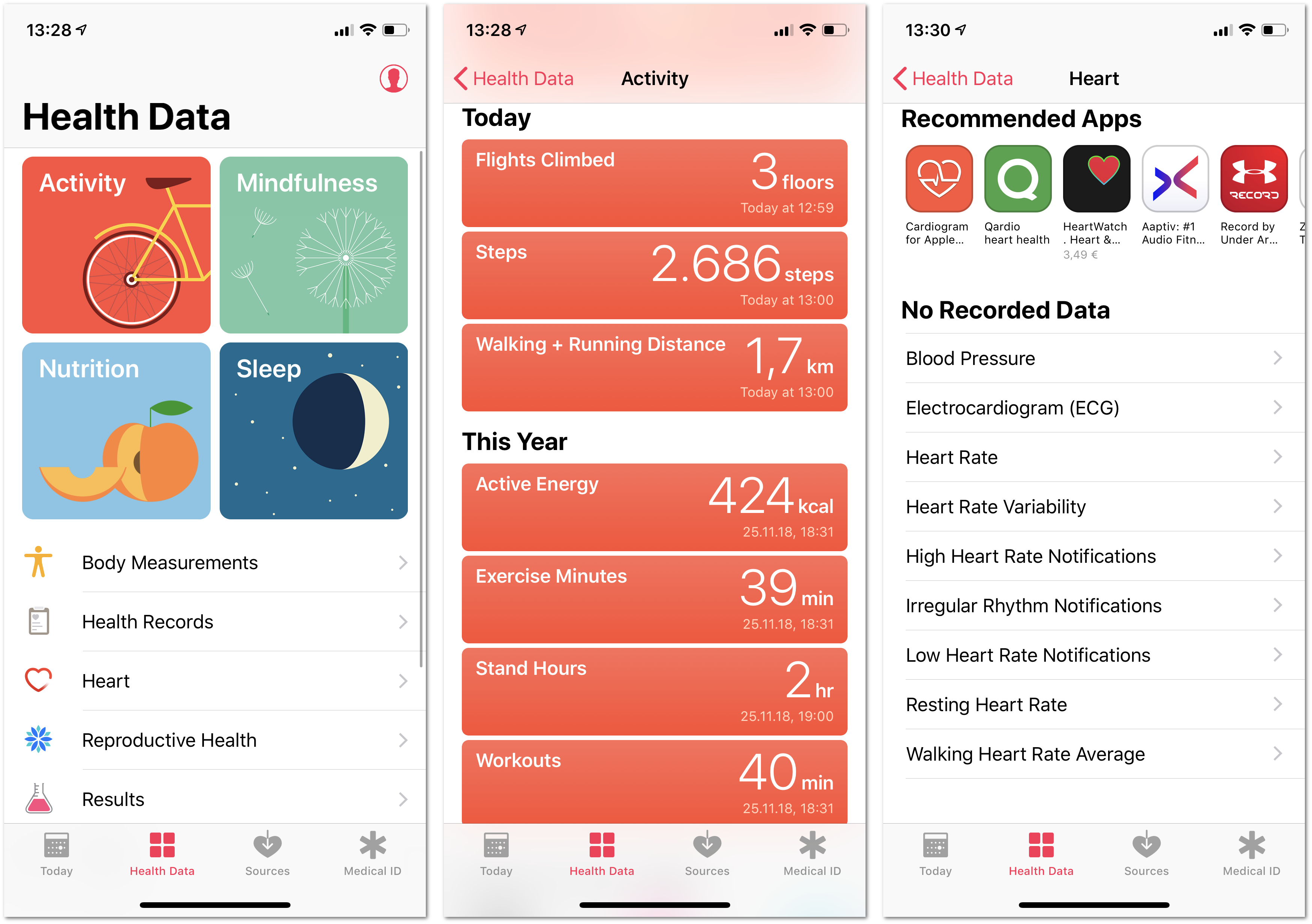}
	\caption{Apple Health Mobile Application}
	\label{fig:health}
\end{figure}

The feasibility of our approach is, thus, highly dependent on the real-time availability of accurate sensor data.
Fortunately, there are commercial data ecosystems in place, such as Apple Health\footnote{https://apple.com/lae/ios/health/} or Samsung Health\footnote{\url{https://www.samsung.com/global/galaxy/apps/samsung-health/}}.
Using these apps on the mobile phone, a user can organize various health data at a single place, which can be visualized and analyzed.
Figure~\ref{fig:health} shows three screenshots of Apple Health.
The first is the overview page, which mentions broad categories like \emph{`Activity'}, \emph{`Mindfulness',} \emph{`Nutrition'}, and \emph{`Sleep'}, but also includes further points like heath-related data and even \emph{`Reproducible Health'}, which is aimed exclusively for women to keep track of their menstrual cycle.
The second screenshots depicts an aggregation of the activities recorded by the iPhone, while the third screenshot shows which heart-related data can be tracked.
It can be seen as a manifestation of the \emph{`Quantified Self'} movement, where the users continuously record their own body data to analyze them with the motif of self improvement~\cite{Swan2013}.
The data is collected via connected wearable devices, entered manually, or by granting read and write access to third party applications on the iPhone as can be seen in the third screenshot.
This constitutes a data ecosystem, which can be used to develop novel applications leveraging this data given the user consents.
Another example is the work of Pandian et~al.~\cite{Pandian2008SmartVW}, a wearable sensor monitoring system that can measure and transmit physiological parameters to a remote monitoring station along with the geo-location of the wearer.

\subsection{Inferring Physiological Conditions}\label{RA1}

\begin{figure*}[t]
	\includegraphics[width=\linewidth]{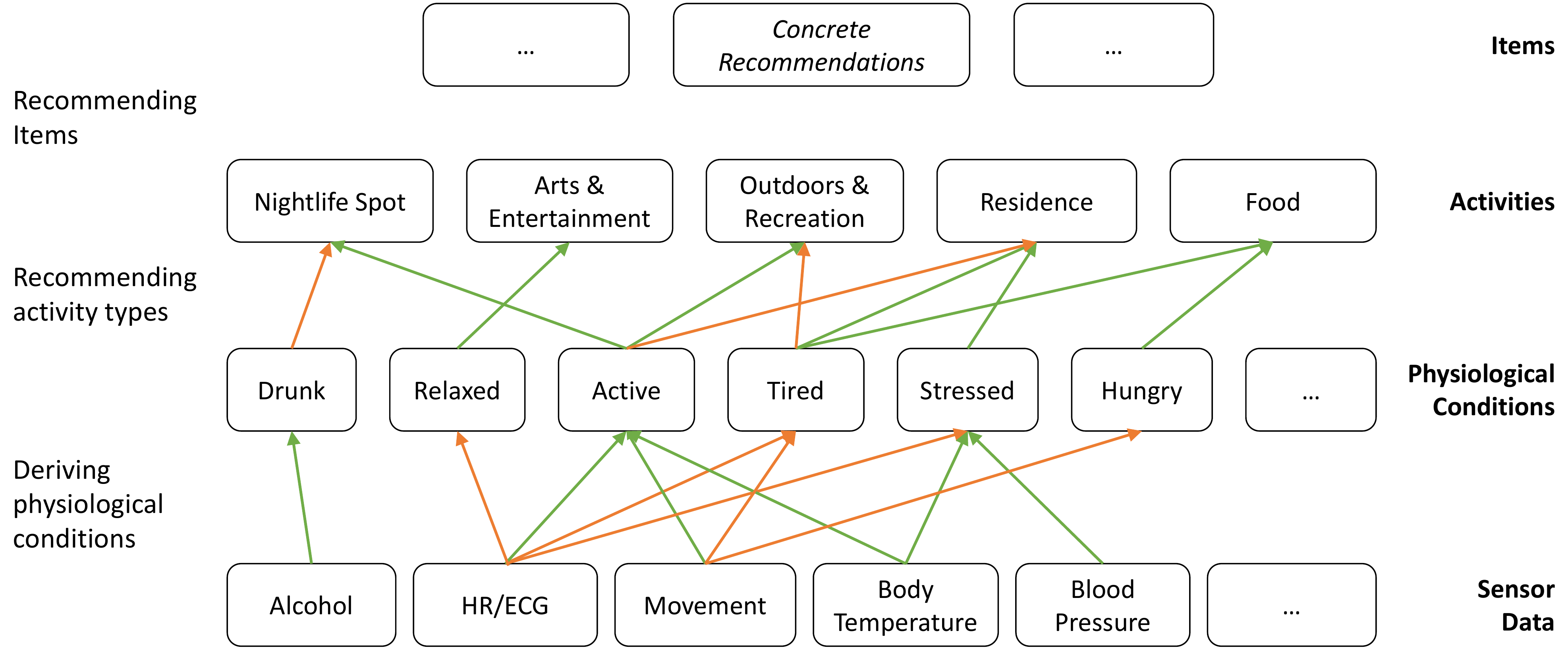}
	\caption{Model of Deriving Physiological Conditions from Sensors and Recommending  Tourist Activities}
	\label{fig:model}
\end{figure*}

Health parameters collected by the wrist-wearable devices have already been  studied and used to deduce relevant physiological conditions from them~\cite{journals/sensors/KamisalicFTK18}.
For example, Alfeo et~al.~\cite{journals/puc/AlfeoBCRPV18} collected information about heart rate and arm motion using wearable devices to assess per-night sleep quality of the wearer.
Zhang et~al.~\cite{journals/inffus/ZhangSA18} used wrist-worn wearable sensors to detect gestures and showed a correlation between the feeding gesture counts and caloric intake more of which lead to overeating. A wrist-worn sensing system was developed by Sugimoto et~al.~\cite{Sugimoto2005} for calculating energy expenditure, by estimating oxygen uptake from correlation between heart rate and oxygen uptake. Defining some thresholds, we can find out if a person is active, relaxed or tired. We chose the following conditions that can be detected using wearable sensor data and are also relevant for travelers.

\begin{itemize}
	\item Active~\cite{Sugimoto2005, journals/sensors/KamisalicFTK18}
	\item Relaxed~\cite{Sugimoto2005, journals/sensors/KamisalicFTK18}
	\item Tired~\cite{Sugimoto2005, journals/sensors/KamisalicFTK18, journals/puc/AlfeoBCRPV18}
	\item Drunk~\cite{Panneer_Selvam_2016}
	\item Hungry~\cite{journals/inffus/ZhangSA18, Sugimoto2005}
	\item Stressed~\cite{journals/jbi/GjoreskiLGG17, conf/iwinac/SandulescuAEBM15, journals/sensors/KamisalicFTK18}
\end{itemize}

\section{Deriving Activities from Physiological Conditions}\label{RA2}

The physiological condition of a tourist influences the suitability of the recommended activities.
Literature lists various activity taxonomies~\cite{Mehmetoglu2007}, however, in this research we fall back to a subset of Foursquare's venue categories\footnote{https://developer.foursquare.com/docs/resources/categories}, as  tourist activities for our data model.
The main advantage is that they cover most touristic activities and provide a direct mapping to concrete items, which is useful for the final recommendation step.
The five categories are
\begin{inparaenum}
\item {Outdoors \& Recreation},
\item {Arts \& Entertainment},
\item {Food},
\item {Residence}, and
\item {Nightlife}.
\end{inparaenum}

Which activity should be suggested to the tourist depends on her current contextual and physiological conditions. For example:

\begin{itemize}
	\item If the tourist is very active, she can be suggested to do sports or adventurous outdoor activities like horse-riding.
	\item If a traveler has not slept for a long time or is stressed, recommending a relaxing activity might be  fitting.
	\item Contrary, if she is relaxed, we would suggest her an intellectual activity.
	\item Finally, if someone has been very active over a prolonged time without taking a break, we would recommend to refresh herself with food or drinks.
\end{itemize}

The above examples describe assumptions of one might expect tourists to behave in general under their particular physiological conditions.
In reality, people might not follow that and act more peculiarly.
Therefore, a context-aware recommender system need to learn the weights between the physiological conditions derived from wearable sensors and the respective activities.
For this, a machine learning algorithm shall be employed to observe the behavior of the tourists given their physiological conditions.
The upper part of \autoref{fig:model}, depicts selected positive and negative influences from physiological conditions to tourist activities.
Green arrows imply a positive correlation, i.e., if the condition has a higher value, the activity at the end of the arrow is recommended.
If the arrow is orange, the score of corresponding activity is decreased.
In our proposed model, the user model comprises a normalized value of each physiological conditions.

The associations between the sensor values and the physiological conditions in the lower part of \autoref{fig:model} can be derived as discussed in \autoref{RA1}.
Then, we can form a 1X6 physiological condition vector, $PC$ (see \autoref{eq:eqPC}), that denotes the current status of a tourist's physiological conditions, each column having a value, $c_{a}, c_{r}, c_{t}, c_{d}, c_{h}, c_{s}$ for each of the six physiological conditions: \underline{a}ctive, \underline{r}elaxed, \underline{t}ired, \underline{d}runk, \underline{h}ungry and \underline{s}tressed.

\begin{equation}\label{eq:eqPC}
  \begin{array}{l@{{}={}}c}
  \text{PC} & \left(\begin{array}{@{}cccccc@{}}
    c_{a}, & c_{r}, & c_{t}, & c_{d}, & c_{h},& c_{s} \\
  \end{array}\right)
  \end{array}
\end{equation}

Next, we use the previously learned associative weights of our activity variables, $or$ (outdoors \& recreation), $ae$ (arts \& entertainment), $fd$ (food), $rs$ (residence), and $nl$ (nightlife) to compute the fitting activity.
This is done by forming a $6X5$ weight matrix, $W$, as shown in \autoref{eq:eqW}.
Each column holds the values yielded from the machine learning for each activity, under each of the physiological condition in each row.
For example, $w_{(or/a)}$ denotes the weight value of opting for activities from the group `outdoor and recreation' having an `active' physiological condition.

\begin{equation}\label{eq:eqW}
  \begin{array}{l@{{}={}}c}
  \text{W} & \left(\begin{array}{@{}cccccc@{}}
    w_{(or|a)} &w_{(ae|a)}&w_{(fd|a)}&w_{(rs|a)}& w_{(nl|a)}\\
    w_{(or|r)} &w_{(ae|r)}&w_{(fd|r)}&w_{(rs|r)} &w_{(n|r)} \\
    w_{(or|t)} &w_{(ae|t)}&w_{(fd|t)}&w_{(rs|t)}& w_{(nl|t)}\\
    w_{(or|d)} &w_{(ae|d)}&w_{(fd|d)}&w_{(rs|d)}& w_{(nl|d)}\\
    w_{(or|h)} &w_{(ae|h)}&w_{(fd|h)}&w_{(rs|h)}& w_{(nl|h)}\\
    w_{(or|s)} &w_{(ae|s)}&w_{(fd|s)}&w_{(rs|s)}& w_{(n|s)}\\
  \end{array}\right)
  \end{array}
\end{equation}

$PC . W$ results in the $1X5$ Activity Recommendation Index  (ARI)  vector as shown in \autoref{eq:eqR}.

\begin{equation}\label{eq:eqR}
  \begin{array}{l@{{}={}}c}
  \text{ARI = PC . W} & \left(\begin{array}{@{}ccccc@{}}
    a_{or}, &a_{ae},&a_{fd},&a_{rs},& a_{nl} \\
  \end{array}\right)
  \end{array}
\end{equation}

The tourist activity to be recommended is the one with the highest ARI value.
Now the recommender system will choose the an item from this category based on the user preferences.

\section{Conclusions and Future Work}\label{CAndFW}

This paper discusses a method to use wearable devices for deriving the physiological  context of a tourist and by this to improve proactive recommendations of activities during a trip.
The sensors equipped in the wearables monitor various physiological parameters of a tourist, and by further processing this data, it is possible to derive the physiological context of a traveler such as being hungry or stressed.
The main contribution is a model to facilitate the already available data from health-related application such as Apple Health to recommend tourist activities.

In future, we plan to implement and evaluate the model in a mobile tourist recommender system.
The challenge is to convince users to make their health-related data available for the recommender for potentially more fitting recommendations.
Since data ecosystems like Apple Health make it easy to share it, we are optimistic that this is possible in general.
We hope  that the users' trust can be achieved by open-sourcing the application and keeping the data on the device of the user.

\bibliographystyle{apalike}
\bibliography{references.bib}

\begin{thebibliography}{}

\bibitem[Adomavicius et~al., 2011]{journals/aim/AdomaviciusMRT11}
Adomavicius, G., Mobasher, B., Ricci, F., and Tuzhilin, A. (2011).
\newblock Context-aware recommender systems.
\newblock {\em AI Magazine}, 32(3):67--80.

\bibitem[Alfeo et~al., 2018]{journals/puc/AlfeoBCRPV18}
Alfeo, A.~L., Barsocchi, P., Cimino, M. G. C.~A., Rosa, D.~L., Palumbo, F., and
  Vaglini, G. (2018).
\newblock Sleep behavior assessment via smartwatch and stigmergic receptive
  fields.
\newblock {\em Personal and Ubiquitous Computing}, 22(2):227--243.

\bibitem[Ashley-Dejo et~al., 2016]{AshleyDejo2016ACP}
Ashley-Dejo, E., Ngwira, S.~M., and Zuva, T. (2016).
\newblock A context-aware proactive recommender system for tourist.
\newblock {\em International Conference on Advances in Computing and
  Communication Engineering}, pages 271--275.

\bibitem[Atembe, 2015]{RolandAtembe2015}
Atembe, R. (2015).
\newblock The use of smart technology in tourism: Evidence from wearable
  devices.
\newblock {\em Journal of Tourism and Hospitality Management}, 3(6):224--234.

\bibitem[Baltrunas et~al., 2012]{Baltrunas2012}
Baltrunas, L., Ludwig, B., Peer, S., and Ricci, F. (2012).
\newblock Context relevance assessment and exploitation in mobile recommender
  systems.
\newblock {\em Personal and Ubiquitous Computing}, 16(5):507--526.

\bibitem[Billinghurst and Starner, 1999]{journals/computer/BillinghurstS99}
Billinghurst, M. and Starner, T. (1999).
\newblock Wearable devices: New ways to manage information.
\newblock {\em Computer}, 32(1):57--64.

\bibitem[Coyle et~al., 2010]{Coyle2010BIOTEXBiosensingTF}
Coyle, S., Lau, K.~T., Moyna, N., O'Gorman, D., Diamond, D., Francesco, F.~D.,
  Costanzo, D., Salvo, P., Trivella, M.~G., Rossi, D.~D., Taccini, N.,
  Paradiso, R., Porchet, J.-A., Ridolfi, A., Luprano, J., Chuzel, C., Lanier,
  T., Revol-Cavalier, F., Schoumacker, S., Mourier, V., Chartier, I., Convert,
  R., De-Moncuit, H., and Bini, C. (2010).
\newblock Biotex—biosensing textiles for personalised healthcare management.
\newblock {\em IEEE Transactions on Information Technology in Biomedicine},
  14:364--370.

\bibitem[Dey, 2001]{Dey:2001:UUC:593570.593572}
Dey, A.~K. (2001).
\newblock Understanding and using context.
\newblock {\em Personal and Ubiquitous Computing}, 5(1):4--7.

\bibitem[Düking et~al., 2016]{10.3389/fphys.2016.00071}
Düking, P., Hotho, A., Fuss, F.~K., Holmberg, H.-C., and Sperlich, B. (2016).
\newblock Comparison of non-invasive individual monitoring of the training and
  health of athletes with commercially available wearable technologies.
\newblock {\em Frontiers in Physiology}, 7(71).

\bibitem[Gjoreski et~al., 2017]{journals/jbi/GjoreskiLGG17}
Gjoreski, M., Lustrek, M., Gams, M., and Gjoreski, H. (2017).
\newblock Monitoring stress with a wrist device using context.
\newblock {\em Journal of Biomedical Informatics}, 73:159--170.

\bibitem[Godfrey et~al., 2018]{Godfreya2018FromAT}
Godfrey, A., Hetherington, V., Shum, H., Bonato, P., Lovell, N.~H., and Stuart,
  S. (2018).
\newblock {From A to Z: Wearable technology explained}.
\newblock {\em Maturitas}, 113(April):40--47.

\bibitem[Islam and Mukhopadhayay, 2017]{10.1088/978-0-7503-1505-0ch1}
Islam, T. and Mukhopadhayay, S.~C. (2017).
\newblock Wearable sensors for physiological parameters measurement: physics,
  characteristics, design and applications.
\newblock In {\em Wearable Sensors}, 2053-2563, pages 1--31. IOP Publishing.

\bibitem[Kamišalić et~al., 2018]{journals/sensors/KamisalicFTK18}
Kamišalić, A., Fister, I., Turkanović, M., and Karakatič, S. (2018).
\newblock Sensors and functionalities of non-invasive wrist-wearable devices: A
  review.
\newblock {\em Sensors}, 18(6).

\bibitem[Laß et~al., 2017]{Lass2017}
Laß, C., Herzog, D., and Wörndl, W. (2017).
\newblock Context-aware tourist trip recommendations.
\newblock In {\em Proceedings of the 2nd RecSys Workshop on Recommenders in
  Tourism}, pages 18--25.

\bibitem[Meehan et~al., 2013]{conf/percom/MeehanLCM13}
Meehan, K., Lunney, T., Curran, K., and McCaughey, A. (2013).
\newblock Context-aware intelligent recommendation system for tourism.
\newblock In {\em International Conference on Pervasive Computing and
  Communications Workshops}, pages 328--331. IEEE.

\bibitem[Mehmetoglu, 2007]{Mehmetoglu2007}
Mehmetoglu, M. (2007).
\newblock Nature-based tourists: The relationship between their trip
  expenditures and activities.
\newblock {\em Journal of Sustainable Tourism}, 15(2):200--215.

\bibitem[Pandian et~al., 2008]{Pandian2008SmartVW}
Pandian, P.~S., Mohanavelu, K., Safeer, K.~P., Kotresh, T.~M., Shakunthala, D.
  T.~I., Gopal, P., and Padaki, V.~C. (2008).
\newblock Smart vest: wearable multi-parameter remote physiological monitoring
  system.
\newblock {\em Medical Engineering \& Physics}, 30 4:466--77.

\bibitem[Sandulescu et~al., 2015]{conf/iwinac/SandulescuAEBM15}
Sandulescu, V., Andrews, S., Ellis, D., Bellotto, N., and Mozos, O.~M. (2015).
\newblock Stress detection using wearable physiological sensors.
\newblock In Ferr{\'a}ndez~Vicente, J.~M., {\'A}lvarez-S{\'a}nchez, J.~R.,
  de~la Paz~L{\'o}pez, F., Toledo-Moreo, F.~J., and Adeli, H., editors, {\em
  Artificial Computation in Biology and Medicine}, pages 526--532, Cham.
  Springer.

\bibitem[Selvam et~al., 2016]{Panneer_Selvam_2016}
Selvam, A.~P., Muthukumar, S., Kamakoti, V., and Prasad, S. (2016).
\newblock A wearable biochemical sensor for monitoring alcohol consumption
  lifestyle through ethyl glucuronide ({EtG}) detection in human sweat.
\newblock {\em Scientific Reports}, 6(1).

\bibitem[Seneviratne et~al., 2017]{journals/comsur/SeneviratneHNLK17}
Seneviratne, S., Hu, Y., Nguyen, T., Lan, G., Khalifa, S., Thilakarathna, K.,
  Hassan, M., and Seneviratne, A. (2017).
\newblock A survey of wearable devices and challenges.
\newblock {\em IEEE Communications Surveys and Tutorials}, 19(4):2573--2620.

\bibitem[Shehab et~al., 2017]{conf/aisi/ShehabIOEE17}
Shehab, A., Ismail, A., Osman, L., Elhoseny, M., and El-Henawy, I.~M. (2017).
\newblock Quantified self using {IoT} wearable devices.
\newblock In Hassanien, A.~E., Shaalan, K.~F., Gaber, T., and Tolba, M.~F.,
  editors, {\em Proceedings of the International Conference on Advanced
  Intelligent Systems and Informatics 2017}, volume 639, pages 820--831.
  Springer.

\bibitem[Stoppa and Chiolerio, 2014]{journals/sensors/StoppaC14}
Stoppa, M. and Chiolerio, A. (2014).
\newblock Wearable electronics and smart textiles: A critical review.
\newblock {\em Sensors}, 14(7):11957--11992.

\bibitem[Sugimoto et~al., 2005]{Sugimoto2005}
Sugimoto, C., Ariesanto, H., Hosaka, H., Sasaki, K., Yamauchi, N., and Itao, K.
  (2005).
\newblock {Development of a wrist-worn calorie monitoring system using
  bluetooth}.
\newblock {\em Microsystem Technologies}, 11(8-10):1028--1033.

\bibitem[Swan, 2013]{Swan2013}
Swan, M. (2013).
\newblock The quantified self: Fundamental disruption in big data science and
  biological discovery.
\newblock {\em Big Data}, 1(2):85--99.

\bibitem[Tussyadiah, 2013]{Tussyadiah2014}
Tussyadiah, I. (2013).
\newblock Expectation of travel experiences with wearable computing devices.
\newblock In {\em Information and Communication Technologies in Tourism}, pages
  539--552. Springer.

\bibitem[Vlahakis et~al., 2002]{conf/iswc/VlahakisKTIS02}
Vlahakis, V., Karigiannis, J., Tsotros, M., Ioannidis, N., and Stricker, D.
  (2002).
\newblock Personalized augmented reality touring of archaeological sites with
  wearable and mobile computers.
\newblock In {\em Proceedings of the Sixth International Symposium on Wearable
  Computers}, pages 15--22. IEEE.

\bibitem[Wallace and Press, 2003]{Wallace2003}
Wallace, J. and Press, M. (2003).
\newblock Craft knowledge for the digital age.
\newblock In {\em Proceedings of the Sixth Asian Design Conference}, pages
  14--17.

\bibitem[Woerndl and Groh, 2007]{Woerndl2007}
Woerndl, W. and Groh, G. (2007).
\newblock Utilizing physical and social context to improve recommender systems.
\newblock In {\em Proceedings of the 2007 IEEE/WIC/ACM International
  Conferences on Web Intelligence and Intelligent Agent Technology -
  Workshops}, WI-IATW '07, pages 123--128, Washington, DC, USA. IEEE Computer
  Society.

\bibitem[Woerndl et~al., 2011]{Woerndl2011}
Woerndl, W., Huebner, J., Bader, R., and Gallego-Vico, D. (2011).
\newblock A model for proactivity in mobile, context-aware recommender systems.
\newblock In {\em Proceedings of the Fifth ACM Conference on Recommender
  Systems}, RecSys '11, pages 273--276, New York, NY, USA. ACM.

\bibitem[Zhang et~al., 2018]{journals/inffus/ZhangSA18}
Zhang, S., Stogin, W., and Alshurafa, N. (2018).
\newblock I sense overeating: Motif-based machine learning framework to detect
  overeating using wrist-worn sensing.
\newblock {\em Information Fusion}, 41:37--47.

\end{thebibliography}
\end{document}